\documentclass[prb,preprint,draft,amsmath,showpacs]{revtex4}

\usepackage{graphics}\input epsf
\begin{document}
%%%%%%%%%%%%%%%%%%%%%%%%%%%%%%%%%%%%%%%%%%
\title {Spin gap formation in the quantum spin systems TiOX, X=Cl and Br}

\author {P. Lemmens$^1$, K.Y. Choi$^{2}$, R. Valent\'\i$^{3}$, T.
Saha-Dasgupta$^{4}$, E. Abel$^5$, Y.S. Lee$^5$, and F.C. Chou$^6$}

\affiliation{$^1$ Max Planck Inst. for Solid State Research, D-70569 Stuttgart, and
Inst. for Physics of Condensed Matter, Technical University of Braunschweig, D-38106
Braunschweig, Germany}

\affiliation{$^2$ Inst. for Materials Research, Tohoku University, Sendai 980-8577,
Japan}

\affiliation{$^3$ Inst. f\"ur Theoretische Physik, Universit\"at Frankfurt, D-60054
Frankfurt, Germany}

\affiliation{$^4$ S.N. Bose National Centre for Basic Sciences, Salt Lake City, Kolkata
700098, India}

\affiliation{$^5$ Dept. of Physics, MIT, Cambridge, MA 02139, USA}

\affiliation{$^6$ Center for Materials Science and Engineering, MIT,
Cambridge, MA 02139, USA}

\date{\today}

\begin{abstract}
In the layered quantum spin systems TiOCl and TiOBr the magnetic susceptibility shows a
very weak temperature dependence at high temperatures and transition-induced phenomena
at low temperatures. There is a clear connection of the observed transition temperatures
to the distortion of the octahedra and the layer separation. Band structure calculations
point to a relation of the local coordinations and the dimensionality of the magnetic
properties. While from magnetic Raman scattering only a small decrease of the magnetic
exchange by -5-10~\% is derived comparing TiOCl with TiOBr, the temperature dependence
of the magnetic susceptibility favors a much bigger change.
\end{abstract}
\pacs{78.30.-j, 75.10.Jm, 75.30.Et} \maketitle

\section{Introduction}

The compounds TiOX, with X=Cl and Br are formed by layers of distorted $\rm TiO_4X_2$
octahedra. Quantum magnetism in these systems is based on the $\rm Ti^{3+}$ ions with
one electron ($\rm 3d^1$, s=1/2) in a $\rm t_{2g}$ state. The distortion of the
octahedra leads to the predominant occupation of $\rm d_{xy}$ orbitals that form
chain-like direct exchange paths of orbitals along the crystallographic \emph{b} axis of
the compound. In several experiments, as magnetic susceptibility
\cite{beyon93,seidel03}, NMR \cite{imai03}, ESR \cite{kataev03}, X-ray scattering
\cite{lee03,shaz04}, Raman scattering and optical spectroscopy \cite{lemmens03,caimi04},
strong fluctuations and multiple transitions are observed that are attributed to this
spin/orbital system with low dimensionality. So far, with the exception of one IR
investigation \cite{caimi04b}, only the system TiOCl has been investigated thoroughly.
The isostructural TiOBr has not been in the center of interest due to more severe
problems in growing single crystalline samples of sufficient quality. The scaling of the
IR active phonon frequencies with the involved ionic masses, however, proposes TiOBr as
a perfect reference system. The compounds differ crystallographically by an increase of
the \emph{c} axis lattice parameter from 8.03 to 8.53$\AA$, going from X=Cl to Br. This
means that in TiOBr the coupling between the planes of octahedra is even less important
compared to TiOCl. Furthermore, the distortion of the octahedra is larger in TiOBr.

The compounds TiOX belong to a new class of spin-1/2 transition metal oxides based on
$\rm Ti^{3+}$ ions in weakly interconnected, distorted octahedral coordinations
\cite{axtell97,isobe02,seidel03,isobe02b}. Very often these systems show phase
transitions into singlet ground states that resemble to the spin-Peierls instability
\cite{bray83}. In contrast to, e.g. $\rm CuGeO_3$ \cite{hase93,lemmens-rev} where the
spin-Peierls transition leads to a mean-field size of the reduced gap ratio, very large
spin gaps exist in some titanates \cite{isobe02,lemmens-millet}. For TiOCl even a pseudo
gap for T$>$$\rm T_c$ has been reported based on NMR and Raman scattering experiments
\cite{imai03,lemmens03}. As $\rm t_{2g}$ states of the $\rm Ti^{3+}$ ions ($\rm 3d^1$,
s=1/2) in perfect octahedral surrounding show orbital degeneracy, it is tempting to
assign part of these phenomena to orbital degrees of freedom. Spin-orbital coupled
systems have been investigated theoretically with respect to their instabilities
\cite{pati98,yamashita00,koleshuk01,hikihara04} and as they are candidates for exotic
electronic configurations \cite{beyon93,seidel03,hikihara04}. These theoretical
scenarios did not explicitly consider static or dynamic phonon degrees of freedom. For
2D systems with low symmetry exchange paths it is known that phonons stabilize spin
liquids and shift phase lines \cite{starykh96}. Furthermore, orbital configurations
strongly couple to the lattice\cite{valenti03}. Therefore, changes of lattice parameters
are expected to affect the magnetic properties and related instabilities of such systems
considerably.

In this article we present a comparative study of TiOCl and TiOBr using magnetic
susceptibility, Raman scattering and band structure calculations to relate structural
and electronic properties of these systems and to achieve a better understanding of the
observed instabilities. It is shown, that the fluctuation regime in TiOBr is even more
extended compared to TiOCl.

\section{Structural aspects of TiOX}

The layered crystal structure of TiOX is formed by $\rm Ti^{3+}O^{2-}$ bilayers, that
are separated by $\rm Cl^-$ bilayers. The basic $\rm TiCl_2O_4$ units are distorted
octahedra that share edges. These units form double layers in the \emph{ab} plane of an
orthorhombic unit cell with FeOCl-type structure \cite{schaefer58,schaefer67}. In
Fig.~1~(a) a sketch of octahedra within two rows of the double layer is shown. In
Fig.~1~(b), (c) local Ti-O-X coordinations are given in the \emph{bc} and the \emph{ab}
plane, respectively. Important structural parameters and atomic distances can be found
in Table~I to highlight common and emphasize differences induced by substituting Cl and
Br ions. It is evident that the substitution leads to a considerable increase of the
volume of the unit cell. This effect divides up into an enhanced separation of the
layers and to a smaller extend to an elongation of the cell in \emph{b} axis direction.
The octahedra distortion is more pronounced in TiOBr. This is demonstrated in Table~I by
comparing the shift of the Ti ion out of the basal plane of the octahedra along the
\emph{c} axis.

%%%%%%%%%%%%%%%%%%%%%%%%%%
%FIGURE 2
%%%%%%%%%%%%%%%%%%%%%%%%%%
\begin{figure}[t]
      \begin{center}
       \leavevmode
       \epsfxsize=9cm \epsfbox {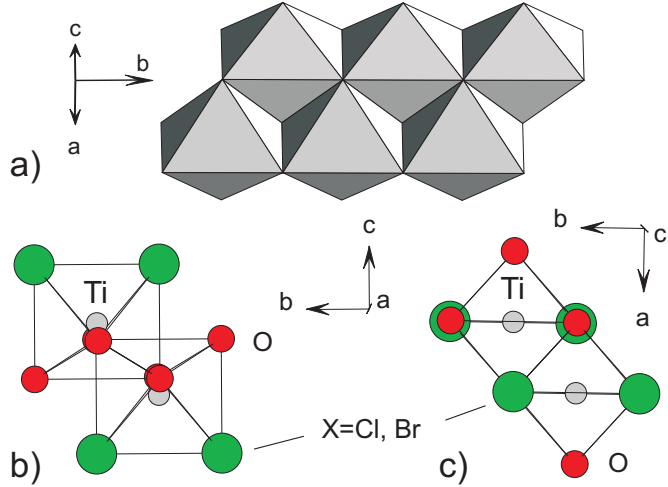}
        \caption{a) A double layer within the crystallographic \emph{ab} plane
        formed by edge sharing, distorted $\rm TiCl_2O_4$ octahedra.
        Local coordinations as projected on the a) bc and c) \emph{ab} plane.
        The relative distances correspond to TiOCl. }
\label{structure}
\end{center}
\end{figure}
%%%%%%%%%%%%%%%%%%%%%%%%%%
%FIGURE 2
%%%%%%%%%%%%%%%%%%%%%%%%%%

\begin{table}[h!]
\caption{Structural parameters and atomic separations for TiOCl and TiOBr with Pmmn (no.
59). Units are $\AA$ or $\AA^3$. Parameters are rounded for clarity and derived using
Refs. \onlinecite{schaefer58} and \onlinecite{schaefer67}. Intra-layer Ti-Ti separations
are given along the direct exchange path parallel to the \emph{b} axis and from the
upper to the lower Ti site of one double layer. For the inter-layer Ti-Ti separations
also two values exist due to the double layers. The distortions of the octahedra are
characterized by $\Delta$c$_{Ti}$, corresponding to the \emph{c} axis shift of the Ti
ion out of its basal plane.} \label{table1}\vspace{0.5cm} \centering
\begin{tabular}{|c|c|c|c|c|}
  \hline
  & lattice parameters & TiOCl & TiOBr & $\rm 10^2$ $\cdot$ $\Delta x/x$\\
  \hline\hline
  unit cell   & a         & 3.79  & 3.79  & $\pm$ 0.0   \\
              & b         & 3.38  & 3.49  & + 3.3 \\
              & c         & 8.03  & 8.53  & + 6.2 \\
  \hline
              & volume    & 102.9 & 112.6 & + 9.4 \\
  \hline\hline
  intra-double layer & Ti - Ti$_b$      & 3.38 & 3.49 & + 3.2 \\
              & Ti - Ti$_{bc}$   & 3.2  & 3.19 & - 0.3 \\
  \hline\hline
  inter-double layer & Ti - Ti$_c$      & 6.58 & 7.12 & + 8.2 \\
              & Ti - Ti$_c$      & 8.13 & 8.91 & + 9.6 \\
  \hline
  distortion  &  $\Delta$c$_{Ti}$& 5.5$\cdot$$\rm 10^{-2}$ & 6.1$\cdot$$\rm 10^{-2}$ & + 10.9 \\
  \hline
\end{tabular}
\end{table}

\section{Sample preparation and Experimental}

Single crystals of TiOCl and TiOBr have been grown using a chemical vapor transport
method.\cite{seidel03} In the following we will describe the preparation of TiOBr as it
is less well established. An initial mixture of $\rm TiO_2$, Ti, and $\rm TiBr_4$ with a
molar ratio of 4:3:9 was sealed in an evacuated quartz tube. The tube was then placed
within a two zone furnace, and a constant thermal gradient was maintained
(650$^{\circ}$C to 550$^{\circ}$C over a 25 cm distance). After approximately 5-8 days,
single crystals of TiOBr with sizes up to 5 $\rm mm^2$ can be extracted. The quality of
the TiOBr crystals is similar to the TiOCl crystals investigated in Ref.
\cite{seidel03}. However, TiOBr is less stable under ambient conditions, as it readily
reacts with water in the air.

The magnetic susceptibility has been measured on a single crystal sample of TiOBr using
a SQUID magnetometer. Raman scattering investigations have been performed using the
514.5~nm excitation wavelength with light polarization parallel to the crystallographic
\emph{b} axis of the platelet-like (ab surface) single crystals of TiOCl
(Ref.~\onlinecite{seidel03}). No strong resonance effects have been detected comparing
514.5~nm and 488~nm excitation wavelength. A comparison of the phonon spectrum of TiOBr
with TiOCl using optical reflectivity $R(\omega)$ experiments can be found in Ref.
\onlinecite{caimi04b}. Details related to optical experiments can be found in
Ref.~\onlinecite{caimi03}. A review on the implications of magnetic Raman scattering in
low dimensional spin systems is given in Ref.~\onlinecite{lemmens-rev}.

\section{Magnetic susceptibility and phase transitions}

%%%%%%%%%%%%%%%%%%%%%%%%%%
%FIGURE 2
%%%%%%%%%%%%%%%%%%%%%%%%%%
\begin{figure}[t]
      \begin{center}
       \leavevmode
       \epsfxsize=9cm \epsfbox {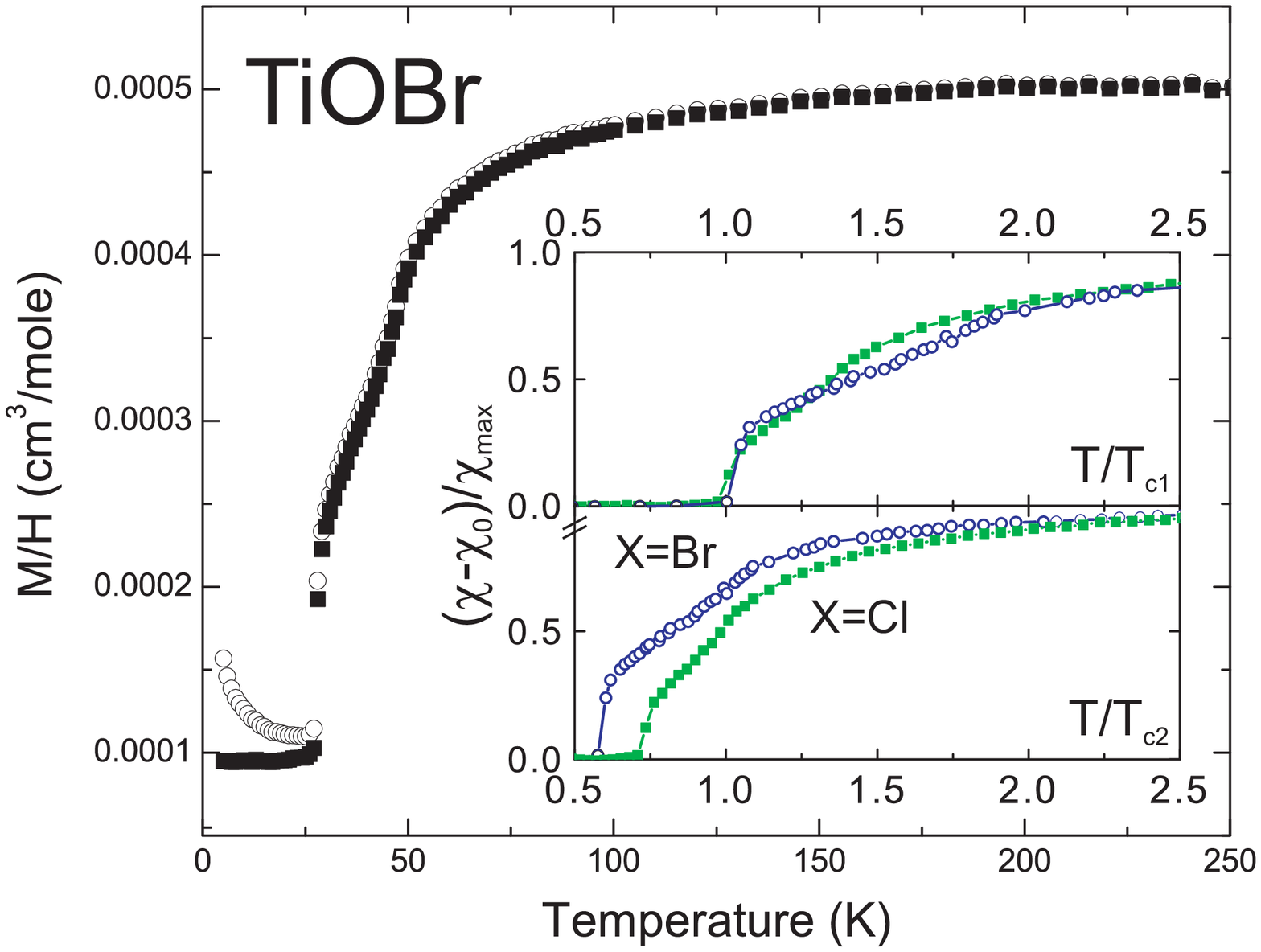}
        \caption{(color online) Magnetic susceptibility (M/H) of TiOBr with
        a magnetic field of 1~Tesla parallel to the \emph{ab} plane.
        Open (full) symbols corresponds to the as measured data
        (data after subtracting a defect contribution). The upper and lower inset
        show the scaled susceptibility ($\chi-\chi_0)/\chi_{max}$ vs. $T/T_{c1}$ or vs. $T/T_{c2}$ for
        TiOX, X=Cl and Br. The transition temperatures are given in Table~II.}
\label{susceptibility}
\end{center}
\end{figure}
%%%%%%%%%%%%%%%%%%%%%%%%%%
%FIGURE 2
%%%%%%%%%%%%%%%%%%%%%%%%%%

In Fig.~2 the magnetic susceptibility $\rm \chi(T)$ of TiOBr is given. The open circles
are the raw data, and the filled squares are the data after subtraction of a small Curie
tail. The general behavior is reminiscent of the susceptibility of TiOCl
\cite{seidel03,imai03} in the sense that above 100~K the magnetic susceptibility $\rm
\chi(T)$ of TiOBr is only weakly temperature dependent \cite{beyon93} and forms a broad
maximum at $\rm T_{max}$$\sim$210~K. In contrast to TiOCl, the Bonner-Fisher curve does
not provide a good fit to the higher temperature behavior of TiOBr (not shown here).

At low temperatures, there is a dramatic drop in the susceptibility at $\rm
T_{c1}^{Br}$=28~K. X-ray scattering in TiOCl has revealed a commensurate dimerization
for T$<$$\rm T_{c1}^{Cl}$=67~K along the \emph{b} axis.\cite{lee03,shaz04} Thereby, the
crystal structure changes from $Pmmn$ to $P2_{1}/m$ with atomic displacements restricted
to the \emph{bc} plane. The flat susceptibility for T$<$$\rm T_{c1}$ is therefore taken
as an indication for a similar structural distortion in TiOBr that accompanies the spin
gap formation. It has been highlighted, that with the exception of the comparably large
spin gap the resulting low temperature state of the two systems is very well comparable
to other spin-Peierls systems \cite{lemmens03,caimi04}.

The inset shows the normalized susceptibility ($\chi-\chi_0)/\chi_{max}$ of both systems
vs. $\rm T_{c1}$ and vs. $\rm T_{c2}$ of the respective compound. $\chi_0$ has been
extrapolated in the limit T $\rightarrow$ 0. From these data it is evident that the
overall susceptibilities scale very well with respect to $\rm T_{c1}$. This refers
mainly to the decrease that is observed towards lower temperatures. Smaller deviations
exist in the temperature regime $\rm T_{c1}$$<$T$<$2$\rm T_{c1}$ and a close inspection
of the susceptibility data reveals a second critical temperature $\rm T_{c2}^{Br}$=47~K
at which the change of susceptibility is most pronounced. At the corresponding
temperature $\rm T_{c2}^{Cl}$=92~K of TiOCl evidence for incommensurate distortions has
been found together with anomalies in the specific heat \cite{lee03}. Some results of
the magnetic characterization are summarized in Table~II.

\begin{table}[h!]
\caption{Magnetic parameters from TiOCl and TiOBr determined from susceptibility, the
maxima positions in Raman scattering data, the exchange coupling parameters from
downfolding the bandstructure and their change in \% with respect to TiOCl.}
\label{table2}\vspace{0.5cm} \centering
\begin{tabular}{|c|c|c|c|}
  \hline
                                   & TiOCl             & TiOBr               & $\rm 10^2$ $\cdot$ $\Delta x/x$\\ \hline\hline
                         $T_{c1}$  & 67~K              & 28~K                & - 58.2 \\ \hline
                         $T_{c2}$  & 92~K              & 47~K                & - 48.9 \\ \hline\hline
            $T_{max}$ in $\chi(T)$ & 400~K             & 210~K               & - 47.5 \\ \hline\hline
               $T_{c1}$/$T_{max}$  & 0.16              & 0.13                & -  \\ \hline
               $T_{c2}$/$T_{max}$  & 0.23              & 0.22                & -  \\ \hline\hline

  $\rm \Delta\omega_{max}^{1}$     & 928~$\rm cm^{-1}$ & 875~$\rm cm^{-1}$   & - 5.7  \\ \hline
  $\rm \Delta\omega_{max}^{2}$     &1404~$\rm cm^{-1}$ & 1390~$\rm cm^{-1}$  & - 1.0  \\
  \hline\hline
  $\rm J_{d-xy}^{downfolding}$       & 621~K             & 406~K               & - 35  \\ \hline
\end{tabular}
\end{table}

In the lower inset of Fig.~\ref{susceptibility} the magnetic susceptibility is plotted
with respect to the higher characteristic temperature. The decrease of the magnetic
susceptibility does not scale well with $\rm T_{c2}$. As this decrease is related to the
opening and magnitude of the spin gap $\Delta(T)$ at the respective temperature, $\rm
T_{c2}$ cannot be directly related to the spin gap formation.

The scaling behavior may be further investigated comparing ratios of the critical
temperatures and the maxima of the susceptibility $\rm \chi(T)$. The maximum in $\rm
\chi(T)$ gives some information about an averaged magnetic coupling strength. According
to Cross and Fisher \cite{cross79} a linear relation between $\rm T_c$ and the exchange
coupling J is expected for spin-Peierls systems with $\rm T_c/J=0.8\lambda$ and a spin
phonon coupling $\lambda$. In Table~II the corresponding ratios are given. It is
noteworthy, that in contrast to the decrease of the susceptibility its maximum position
scales with the higher transition temperature. Comparing these characteristic
temperatures the following ratios are determined: $\rm T_{c1}^{Cl}/T_{c1}^{Br}$ = 2.4,
$\rm T_{c2}^{Cl}/T_{c2}^{Br}$ = 1.96, while the maximum temperatures give $\rm
T_{max}^{Cl}/T_{max}^{Br}$ = 1.91. Especially the latter two ratios show a good matching
and lead to the conclusion that the characteristic energy scales of the thermodynamic
quantities differ roughly by a factor of two between the two systems. This conclusion
does not perfectly match to the calculated hopping integral t given in Table~III and
results from high energy Raman scattering discussed below.

For TiOCl large fluctuations have been reported at high temperatures (T$>$$\rm T_{c1}$)
based on anomalies in the $\rm ^{{47,49}}Ti$ spin relaxation rate and a strong
temperature dependence of the ESR-derived g-factor \cite{imai03,kataev03}. These
observations lead to the assignment of a fluctuation or pseudo gap temperature at
T*=135~K. For the temperature regime $\rm T_{c1}$$<$T$\leq$T*=135~K also pronounced
anomalies and softenings of the Raman-active optical phonon modes exist
\cite{lemmens03}. Furthermore, the spin gap $\rm E_g=430~K$ determined from NMR
\cite{imai03} is very large. The reduced gap ratios are $\rm 2E_g/k_BT_{c1,c2}$ = 10 -
15.

\section{High energy Raman scattering}

%%%%%%%%%%%%%%%%%%%%%%%%%%
%FIGURE 3
%%%%%%%%%%%%%%%%%%%%%%%%%%
\begin{figure}[t]
      \begin{center}
       \leavevmode
       \epsfxsize=12cm \epsfbox {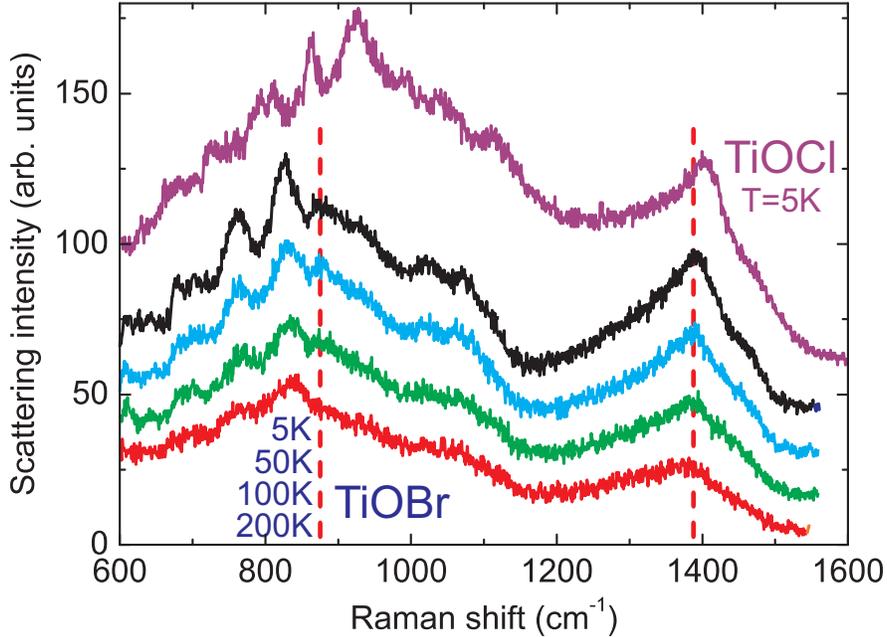}
        \caption{(color online) High energy Raman scattering for TiOX with
        X=Cl and Br with light polarization parallel to the crystallographic $b$ axis, c(bb)-c.
        The data are shifted for clarity. The dashed vertical lines denotes
        the position of the maxima observed in TiOCl.}
        \label{raman}
\end{center}
\end{figure}
%%%%%%%%%%%%%%%%%%%%%%%%%%
%FIGURE 3
%%%%%%%%%%%%%%%%%%%%%%%%%%

In Fig.~3  the Raman scattering intensity of TiOCl (T=5~K) and TiOBr
(5~K$<$T$<$200~K) is given in the frequency regime that is characteristic for magnetic
scattering. Typical energies of two magnetic exchange scattering are estimated within an
Ising-like picture counting the number of broken bonds that an exchange process leaves
behind. Using the coupling constant J=660~K for TiOCl from magnetic susceptibility
\cite{seidel03},  magnetic light scattering should have typical energies between
2$\cdot$J=920~$\rm cm^{-1}$ and 4$\cdot$J=1800~$\rm cm^{-1}$. The exchange pathes
considered theoretically have both one dimensional and two dimensional aspects
\cite{valenti03}. For TiOCl two broad maxima are observed, a symmetric one at 928~$\rm
cm^{-1}$ (2$\cdot$J=917~$\rm cm^{-1}$) and an asymmetric one at 1404~$\rm cm^{-1}$
(3$\cdot$J=1375~$\rm cm^{-1}$). For TiOBr the overall spectral distribution of
scattering intensity has a similar shape. The maxima are at $\rm
\Delta\omega_{max}^{Br-1}$=875~$\rm cm^{-1}$ and $\rm
\Delta\omega_{max}^{Br-2}$=1390~$\rm cm^{-1}$, i.e. they are shifted by 5\% and 1\%,
respectively, to lower frequencies. At low and intermediate energies,  pronounced
oscillations or superstructures are observed with peak frequencies $\rm
\Delta\omega_{max}^{Br-i}$=695, 763, 825 and 883~$\rm cm^{-1}$. These structures have a
mean separation of 63~$\rm cm^{-1}$ and are even more pronounced compared to TiOCl.

%In TiOX, the first more symmetric maximum at approximately (2J), in contrast, resembles
%to the shape of the two-magnon continuum in the spin tetrahedra system $\rm
%Cu_2Te_2O_5Br_2$. In this compound a 2D exchange topology of weakly coupled spin
%tetrahedra is realized that pushes the system to the proximity of a quantum critical
%point \cite{lemmens01}.

The double peak structure with additional modulations at low and intermediate energies
is a complex structure compared to the observations in other 1D or 2D quantum spin
systems. The line shape of such scattering intensity is usually a result of strong
magnon-magnon interactions and the local exchange topology. Magnon-magnon interaction
leads to a characteristic renormalization and broadening of the spectral weight to lower
energy \cite{brenig01}. In undoped high temperature superconductors with $\rm CuO_2$
planes that represent a 2D s=1/2 Heisenberg magnet, a single, broadened peak with a
maximum at $\rm \Delta\omega_{max}^{HTSC}\approx$2.8$\cdot$J is observed
\cite{lyons88,sugai88,sugai99b}. Also in $\rm Cu_2Te_2O_5Br_2$, based on weakly coupled
spin tetrahedra, a single symmetric maximum has been observed at $\rm
\Delta\omega_{max}\approx$2$\cdot$J \cite{lemmens01,gros03,jensen03}. Well-defined
double peak structures have, however, been observed in the 2D nickelates $\rm
La_{2-x}Sr_{x}NiO_{4+d}$ if stripe domains with a modulation of spin/charge exist. The
two maxima are then the result of exchange processes across and within the
antiferromagnetic domain walls and they are related to the exchange coupling as
3$\cdot$J and 4$\cdot$J for a Sr doping of x=1/3 \cite{yamamoto98,blum98,gnezdilov02}.
For TiOCl recently evidence for a more complex structure of dimerized Ti sites has been
obtained from structural investigations \cite{lee03}. It is proposed, that antiphase
domain walls exist within the \emph{ab} planes that separate nano domains of different
dimer orientations. This situations resembles to some extend the situation in the 2D
nickelates and may be the reason for the two maxima.

In the respective frequency range also excitation of other origin can contribute. Here
multi-phonon scattering \cite{choi05b} or magnon-phonon coupled modes \cite{windt01}
should be discussed. Multi-phonon scattering is frequently observed in isolating
transition metal oxides and leads to peak structures with frequencies very close to
multiples of the optical phonon frequencies. For the case of TiOCl the Raman allowed
phonon modes of the high temperature phase consist of in-phase Cl-Ti (203~$\rm cm^{-1}$)
and out-of-phase O-Ti (365~$\rm cm^{-1}$) and Ti-Cl (430~$\rm cm^{-1}$) modes,
respectively. The numbers in brackets give the respective frequencies for TiOCl. These
Raman-active modes and also the IR-active phonon modes considerable soften with the
exchange of Cl by Br following simple scaling relationships \cite{caimi04b}. For TiOBr
phonon Raman modes are observed at 144~$\rm cm^{-1}$, 323~$\rm cm^{-1}$ and 413~$\rm
cm^{-1}$ \cite{choi05}, which correspond to a decrease of the phonon frequencies by
4-29\%. It is also expected that the phonon dispersion is less pronounced in TiOBr due
to the weaker interplane coupling.

In the high energy regime comparing TiOCl with TiOBr no considerable shift is observed,
therefore a dominant contribution of multiphonon scattering to the two maxima is not
suggested. Furthermore, corresponding multiples of the phonon frequencies do not exist
in the high energy Raman spectral range. This statement should be softened with respect
to the modulations seen on the left shoulder of the first more symmetric maximum. It has
been shown for TiOCl that this modulation has the same frequency as the difference of
two optical phonon frequencies \cite{lemmens03}. Finally, we can rule out orbital
scattering to observed double peak structures as the large distortion should lead to
orbital excitations in the range of 0.1-0.3eV. Concluding the Raman scattering results
we state that the corresponding exchange coupling constants do only mildly change with
composition.

\section{ {\it ab initio} Calculations}

%%%%%%%%%%%%%%%%%%%%%%%%%%
\begin{figure}
      \begin{center}
       \leavevmode
       \epsfxsize=8cm \epsfbox {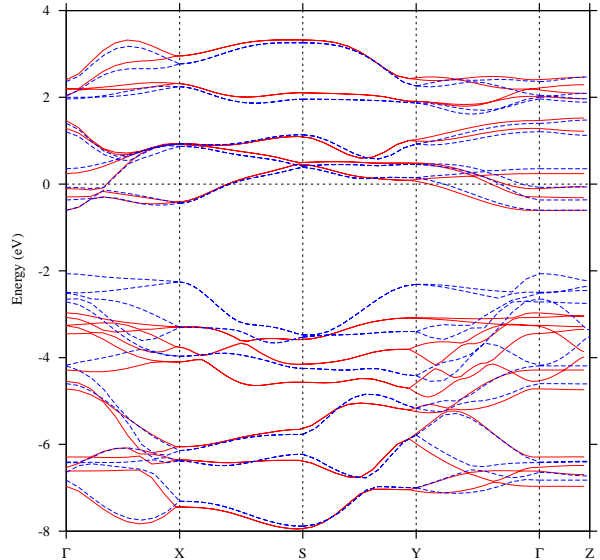}
        \caption{LDA-bandstructure of TiOCl (solid red line) and TiOBr (dashed blue line)
      along the path $\Gamma$= (0, 0, 0),  X= (0, 1/2, 0), S=(-1/2, 1/2, 0), Y=(-1/2, 0, 0), $\Gamma$,
      Z=(0, 0, 1/2)
in units of 2$\pi$/a, 2$\pi$/b, 2$\pi$/c.}
        \label{bands}
\end{center}
\end{figure}
%%%%%%%%%%%%%%%%%%%%%%%%%%

%%%%%%%%%%%%%%%%%%%%%%%%%%
\begin{figure}
      \begin{center}
       \leavevmode
       \epsfxsize=8cm \epsfbox {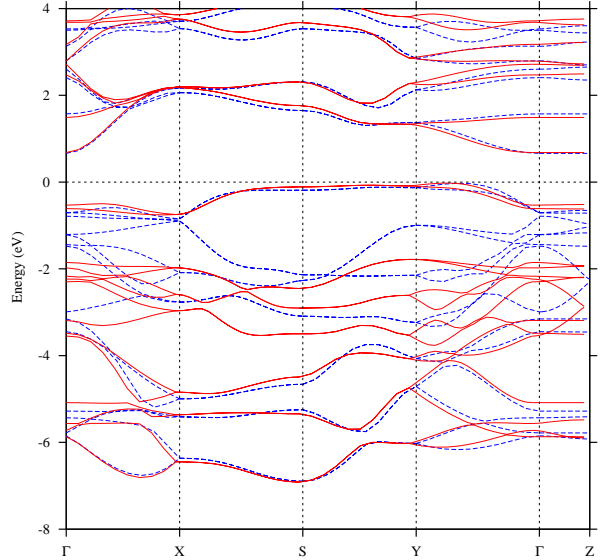}
        \caption{LDA+U bandstructure of TiOCl (solid red line) and TiOBr (dashed blue line)
      along the path $\Gamma$- X - S - Y - $\Gamma$ - Z. Note the opening of the gap at
      the Fermi level with respect to the LDA results. The two bands right below  E$_F$
       are of Ti-$\rm d_{xy}$ nature.}
        \label{bands_U}
\end{center}
\end{figure}
%%%%%%%%%%%%%%%%%%%%%%%%%%

In order to investigate the electronic properties of TiOX, we carried out density
functional calculations for TiOBr  and compared with the results obtained for
TiOCl\cite{seidel03,valenti03}. We performed our {\it ab initio} study in the local
density approximation (LDA), the generalized gradient approximation (GGA)\cite{perdew96}
and in the so-called LDA+U\cite{anisimov97} by using the linearized muffin tin orbital
(LMTO) method based on the Stuttgart TBLMTO-47 code\cite{andersen75}. The results within
LDA and GGA did not lead to significant differences.  An analysis of the TiOBr
bandstructure (see Fig.\ \ref{bands}) shows that the shape of the LDA-$\rm t_{2g}$ bands
crossing the Fermi level have almost identical dispersions to
TiOCl\cite{seidel03,valenti03}, what indicates that the exchange paths are similar
between these two compounds. Nevertheless, there is in TiOBr a narrowing of the
bandwidth along the Y-$\Gamma$-Z path which is a consequence of the enlargement of the
cell in $b$ and $c$ directions.

The calculation within the LDA+U approach shows for TiOBr Ti-$d_{xy}$ to be the
groundstate as in TiOCl (see Fig.\ \ref{bands_U} ). In order to get a reliable estimate
of the interaction paths in TiOBr, we applied the tight-binding-downfolding
procedure\cite{saha-dasgupta00,valenti03} which obtains the effective Ti-$\rm
d_{xy}$-Ti-$\rm d_{xy}$ hopping parameters by downfolding all the degrees of freedom in
the bandstructure calculation other than Ti-$\rm d_{xy}$. The predominant hopping path
in TiOX, $\rm t$, is along the nearest neighbor (n.n.) Ti-$\rm d_{xy}$ in the $b$
direction. $\rm t=-0.21eV$ for TiOCl while $\rm t=-0.17eV$ for the Br system. This
reduction reflects the narrowing of the bands in TiOBr.

\begin{table}
\caption{Predominant hopping integrals (in eV) $t_{i,j}$ for TiOCl and TiOBr obtained
from downfolding the bandstructure results (see text). In the Table only the significant
digits have been given.} \label{parameters}\begin{center}
\begin{tabular}{|c c|c|c|c|}
\hline

          & Ti-Ti hopping integrals                        & TiOCl & TiOBr  & $\rm 10^2$ $\cdot$ $\Delta x/x$\\
          \hline \hline
     t        & n.n. hopping along $b$                         & -0.21 &  -0.17 & - 19 \\
  t$^{'}_{b}$ & n.n.n. hopping along $b$                   & -0.03 &  -0.04 & + 33 \\  \hline\hline
      t$^{'}$ &   n.n. hopping along $a$, same layer           & 0.04  &  0.06  & + 50 \\
     $t^{''}$ & n.n. hopping along \emph{a}, different layers  &  0.03  &  0.04  & + 33 \\
  \hline\hline
 %  & &  &  &  & \\
\end{tabular}
\end{center}

\end{table}

A rough estimate of the antiferromagnetic superexchange along $b$ can be obtained by
using the expression $\rm J=4t^2/U$  which for $\rm U$=3.3eV is J $\approx$ 621~K for
TiOCl and J $\approx$ 406~K for TiOBr. This change is larger than the shift seen in high
energy Raman spectra and in qualitative agreement with the susceptibility measurements.
Since we are here interested in estimating the differences between TiOCl and TiOBr, we
present in Table \ref{parameters} a detailed account of all relevant effective hopping
parameters between Ti-$\rm d_{xy}$-Ti-$\rm d_{xy}$. We observe that the hopping
integrals along other paths than $\rm t$ are almost an order of magnitude smaller than
the main hopping along $b$, nevertheless the TiOBr compound has slightly larger
effective hoppings along the $ab$ plane than TiOCl what indicates that in that system
the interactions within the $ab$ plane may be more significant than in TiOCl. The
comparison (Table~II) shows that this result is only in qualitative agreement with the
trend seen in maximum of the magnetic susceptibility.

\section{Summary and Conclusions}

The system TiOX, with X=Cl and Br, shows a rich spectrum of anomalies related to
electronic, spin and structural degrees of freedom that goes far beyond the scenario of
usual spin-Peierls materials \cite{cross79}. The analysis of the magnetic susceptibility
and phase diagram shows that the low temperature transition, $\rm T_{c1}$, connected
with a commensurate structural distortion, scales with the decrease of the
susceptibility and the evolution of the spin gap. The high temperature transition, $\rm
T_{c2}$, scales with the maximum in the susceptibility that is itself related to the
characteristic energy scale of the magnetic system.

The deviations between the latter energy scale and the high energy Raman scattering are
either due to a complex structure of the dimer formation or due to an interplay with the
phonon sector. Due to the strong distortion and the already weak interlayer coupling in
TiOCl we do not expect that changes of the dimensionality play an essential role.

Finally, our band structure calculations have shown that, although the effective Ti-$\rm
d_{xy}$-Ti-$\rm d_{xy}$ hopping parameters are renormalized with the substitution of Cl
by Br, the electronic spectrum does not change drastically. This is in accordance with
the high energy Raman scattering. Following the Cross-Fisher scaling, the difference in
the transition temperatures of the two compounds should then be attributed both to the
smaller electronic bandwidth and a modified spin-phonon coupling. This scenario,
however, does neither explain the strong fluctuations nor the large spin gap observed
even above the transition temperatures. Further studies on both systems are needed to
elucidate these points.

% Why is Tc so much different? Tmax is a factor 2 different, J may not be that different
% The dimensionality of the spin system is not important as the transition is related to
% the coupling to a phonon system

{\bf Acknowledgement}

The authors acknowledge fruitful discussions with C. Gros, H. Rosner, L. Degiorgi,
E.~Ya. Sherman, R. Kremer and B. Keimer. This work was supported by the MRSEC Program of
the National Science Foundation under award number DMR 02-13282, and DFG SPP1073.

%\bibliography{lit}
            %\bibliographystyle{prsty}

            %\bibliographystyle{peter}
            % print out full name list
            %\addcontentsline{toc}{chapter}{Bibliography}

%\bibitem{Saha-Dasgupta_00} O.K. Andersen and T. Saha-Dasgupta, Phys. Rev. {\bf
%B62}, R16219 (2000) and references therein.

\end{document}